\documentclass[a4paper,fleqn,usenatbib]{mnras}

\usepackage{newtxtext,newtxmath}

\usepackage[T1]{fontenc}
\usepackage{ae,aecompl}

\usepackage{graphicx}	
\usepackage{amsmath}	
\usepackage{amssymb}	

\newcommand{\hei}{\mbox{He{\;}\textsc{i}}}
\newcommand{\heii}{\mbox{He{\;}\textsc{ii}}}
\newcommand{\hefive}{\ensuremath{\text{He}\;\textsc{ii}~\lambda{4541}}}
\newcommand{\henine}{\ensuremath{\text{He}\;\textsc{i}~\lambda{4922}}}
\newcommand{\kms} {\mbox{\rm km$\;$s$^{-1}$}}

\newcommand{\vesini}{\ensuremath{v_{\rm e}\sin{i}}}
\newcommand{\vmsini}{\ensuremath{v_{\rm m}\sin{i}}}
\newcommand{\teff}{\ensuremath{T_{\rm eff}}}
\newcommand{\teffl}{\ensuremath{T^{\ell}_{\rm eff}}}
\newcommand{\logg}{\ensuremath{\log{g}}}
\newcommand{\loggl}{\ensuremath{\log{g}^\ell}}
\newcommand{\loggp}{\ensuremath{\log{g}_{\rm p}}}
\newcommand{\omcrit}{\ensuremath{\omega_{\rm c}}}
\newcommand{\omomc}{\ensuremath{\omega/\omega_{\rm c}}}
\newcommand{\rpole}{\ensuremath{R_{\rm p}}}

\newcommand{\refrsun}{\ensuremath{\mathcal{R}_{\odot}}}

\title[Rotation in O stars]{Are the O stars in WR+O binaries exceptionally rapid rotators?}

\author[Dominic Reeve and Ian D. Howarth]{
Dominic Reeve and
Ian D. Howarth\thanks{E-mail: i.howarth@ucl.ac.uk}
\\
Department of Physics and Astronomy, University College London,
Gower Street, London WC1E 6BT, UK
}

\date{Accepted 2018 May 9. Received 2018 May 9; in original form 2018
  Mar 16.}

\pubyear{2018}

\begin{document}
\label{firstpage}
\pagerange{\pageref{firstpage}--\pageref{lastpage}}
\maketitle

\begin{abstract}
  We examine claims of strong gravity-darkening effects in the O-star
  components of WR+O binaries.  We generate synthetic spectra
  for a wide range of parameters, and show that the line-width results
  are consistent with extensive measure\-ments of O stars that are
  either single or are members of `normal' binaries.  By contrast, the
  WR+O results are at the extremes of, or outside, the distributions
  of both models and other observations.  Remeasure\-ment of the WR+O
  spectra shows that they can be reconciled with other results by
  judicious choice of pseudo-continuum normalization.  With this
  interpretation, the supersynchronous rotation previously noted for
  the O-star components in the WR+O binaries with the longest orbital
  periods appears to be unexceptional.  Our investigation is therefore
  consistent with the aphorism that if the title of a paper ends with
  a question mark, the answer is probably `no'.
\end{abstract}



\section{Introduction}

Rotation is known to be a significant factor in massive-star
evolution, giving rise to internal mixing (\citealt{Eddington25})
which has consequences both for directly observable quantities, such as luminosity
and surface abundances (e.g., \citealt{Sweet53}; \citealt{Heger00}),
and for the stars' lifetimes and ultimate fates (e.g.,
\citealt{Maeder12}; \citealt{Langer12}).  

The most rapid rotators are expected to exhibit gravity darkening: a
reduction in local surface temperature (and hence flux) that is
proportional to local effective gravity \citep{vonZeipel24}, resulting
in the equatorial regions being cooler than the poles.  This
expectation has been substantiated indirectly, through spectroscopy
(e.g., \citealt{Walker91}, \citealt{Howarth93}), and directly, through optical
long-baseline interferometric imaging, which additionally reveals the
distortion in surface shape arising from centrifugal forces (e.g.,
\citealt{DomdeSou03}).

Recently \citeauthor{Shara17} (\citeyear{Shara17}, hereinafter S17)
have published an analysis of good-quality echelle spectroscopy of a
number of Galactic binaries each composed of a Wolf-Rayet (WR) and an
O-type star, with the aim of measuring rotational velocities for the
O-star components.  The challenges of such measure\-ments are
demonstrated by the fact that prior to their study results had been
published for only two such systems; S17 were able to extend the sample
to eight targets.  For all systems investigated, they found the O-star
\hei\ absorption lines to be systematically broader than their \heii\
counterparts, in terms of both directly measured line widths, and
inferred rotational speeds.  They interpreted this result in the
context of strong gravity-darkening effects arising from rapid
rotation, such that \heii\ line formation largely arises in hot
polar caps, while \hei\ lines are formed at equatorial latitudes.

Such rapid rotation would have significant implications for
angular-momentum transfer in massive binary systems, for (orbital)
circularization and (rotational) synchronization, and hence for binary
evolution, as well as having broader ramifications of the
interpretation of rotation in currently, or effectively, single
O~stars. If validated, the S17 inferences would therefore have important
consequences; this alone is sufficient to motivate subjecting them to
further scrutiny.  Additionally, however, there are some apparently
anomalous aspects of their conclusions which prompt caution.  

First
among these is simply the magnitude of the reported effects, reaching
up to a factor $\sim$two difference in apparent projected velocities
for the \hei\ and \heii\ lines.  This is considerably larger than the
$\sim$10\%\ effects predicted for Be stars (e.g.,
\citealt{Townsend04}), or observed in the most rapidly rotating single
Galactic O~stars (e.g., \citealt{Howarth01}).  Furthermore, although
the projected equatorial speeds inferred by S17 are reasonably large,
they are in all cases thought to be substantially subcritical, with
angular rotation rates reported to be typically only $\sim$65\%\
of the critical value at which the effective gravity is zero at the
equator,
\begin{align}
\omcrit = \sqrt{
{(G M_*)}/ {(1.5 \rpole)^3}
}
\label{eq:vcrit}
\end{align}
(for a star of
mass $M_*$ and polar radius \rpole).
Consequently, it is surprising that
dramatic gravity-darkening effects should be manifest in these
systems, when such strong signatures have not been found in
well-studied single stars.

To explore these issues, we have calculated synthetic spectra for a
grid of model rotating stars (Section~\ref{sec:mod}), and compared
these to a range of observations (Section~\ref{sec:res}).  The results
of this comparison are discussed in Section~\ref{sec:disco}, along with
an indication of how the inferred results for the WR+O systems may be
reconciled with expectations.

\section{Models}
\label{sec:mod}

\subsection{Basic assumptions}

The geometry is that of a rotationally distorted (Roche-model) stellar
surface, divided into a large number of `tiles'.  The specific
intensity (or radiance) for each tile is interpolated from a
pre-computed grid of model-atmosphere results, as a function of
wavelength $\lambda$, viewing angle $\mu$,\footnote{Where $\mu
  =\cos\theta$ and $\theta$ is the angle between the surface normal
  and the line of sight.} local effective temperature \teffl, and
local effective gravity $\loggl$, Doppler shifted according to the
line-of-sight velocity.  Results for all tiles are summed, weighted by
projected area, in order to
generate a synthetic spectrum.  The model is described in greater
detail by \citeauthor{Howarth01} (\citeyear{Howarth01}; see also
\citealt{Howarth16}).

The use of specific intensities means that limb darkening is taken
into account in a fully wavelength-dependent manner.  Gravity
darkening is modeled in the `ELR' formalism \citep{Espinosa11}, which
gives results close to traditional von~Zeipel gravity darkening
\citep{vonZeipel24}, but which leads to better agreement with, in
particular, interferometric observations (e.g., \citealt{DomdeSou14}).

The model-atmosphere intensities were computed on a dense wavelength
grid, resolving intrinsic line profiles, by using Hubeny's {\sc synspec} code,\footnote{{\tt
    http://nova.astro.umd.edu/Synspec49/synspec.html}} starting from
the atmospheric structures of the \textsc{tlusty} \textsc{Ostar2002}
and \textsc{Bstar2006} grids (\citealt{Lanz03, Lanz07}); abundances and
  micro\-turbulence parameters were as discussed by \citet{Reeve16}.
  The models are line-blanketed, non-LTE, steady-state,
  plane-parallel, and hydro\-static.  The hydro\-static approximation may
  be questioned for hot, low-gravity atmospheres; \citet{Lanz03}
  address this issue at some length, concluding that {\sc tlusty}
  models give a satisfactory representation of most spectral lines in
  the UV--IR regime, and that line blanketing is the more important
  consideration.  For the most rapid rotators other factors (particularly
  gravity darkening) are likely to dominate.

\begin{figure*}
\includegraphics[scale=0.6,angle=-90]{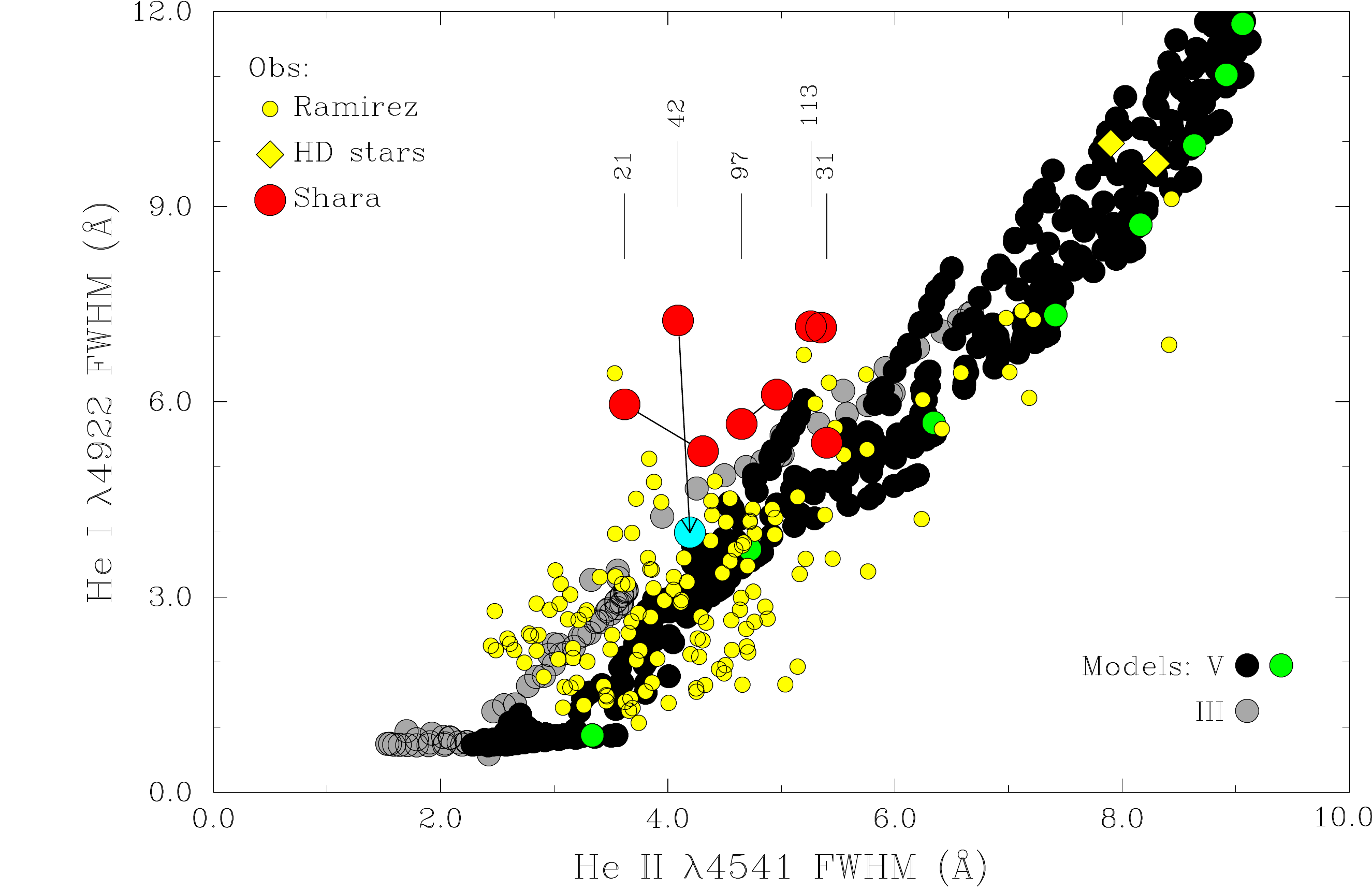}
\caption{Summary of FWHM results for the \henine\
and \hefive\ lines.   Models are as discussed in
$\S$\ref{sec:mod};  green dots identify main-sequence models
at $\teff = 33$~kK, $i = 90^\circ$, to indicate the trends of projected
equatorial rotation velocities ($\vesini = 0$, 141,
240, 313, 368, 410, 443, and 468~\kms).
The `Ramirez' observations  are FWHM measure\-ments 
used by \citet{Ramirez15}, which are unpublished results from
\citet{Sana13}.  `HD' shows new measure\-ments of line widths in 
echelle spectra of the rapid rotators HD~93521 and~149757.  The 
\citet{Shara17} measure\-ments are identified by WR catalogue number
\citep{vanderHucht01}; multiple measure\-ments of the same star are
joined by solid lines, with the leftmost observation labelled.
The arrow indicates the remeasure\-ment of WR~42 discussed in $\S$\ref{sec:disco}.}
\label{fig1}
\end{figure*}

\begin{figure}
\includegraphics[scale=0.6,angle=-90]{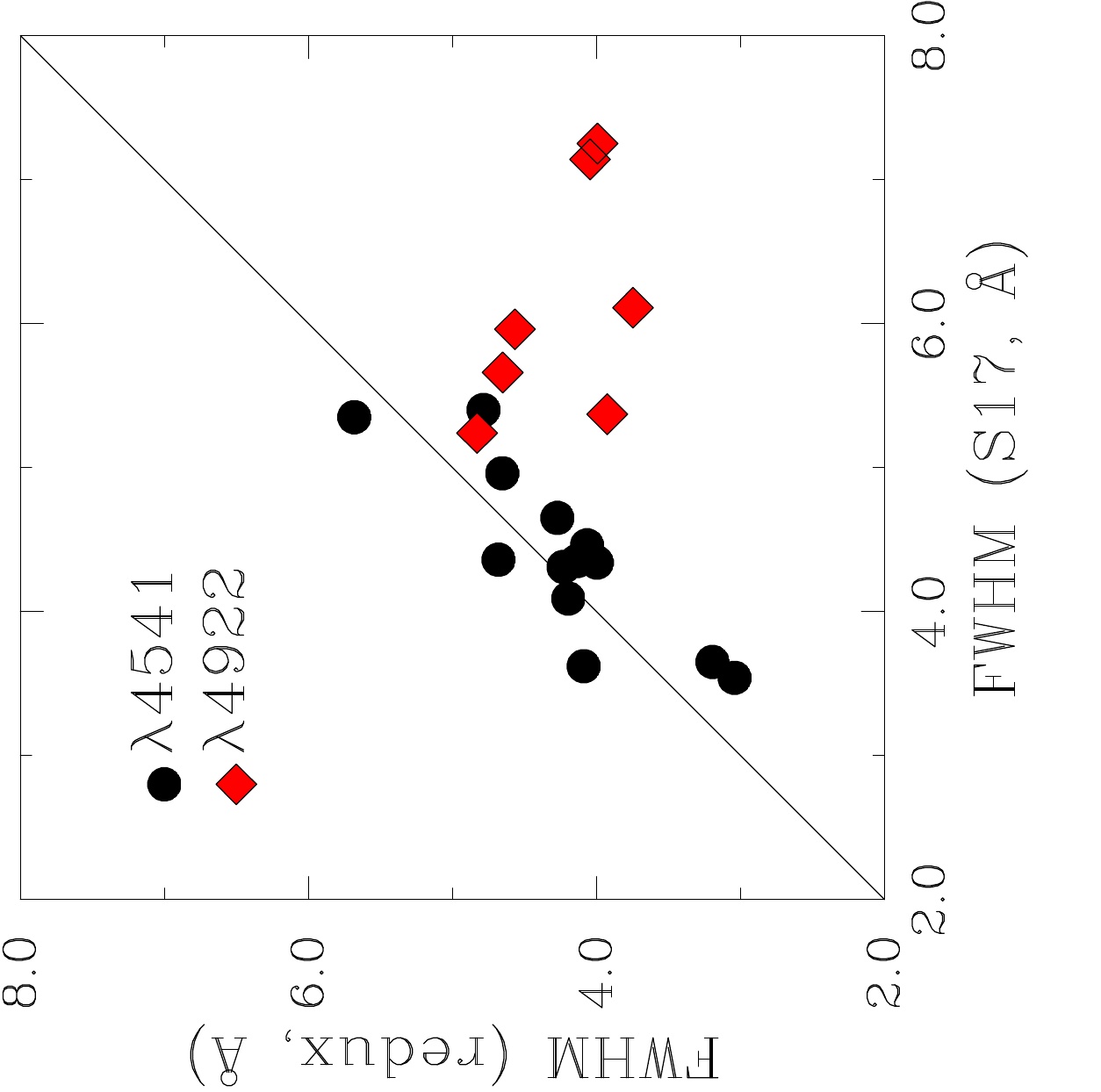}
\caption[FWHM plot2]{Full-widths at half maximum depth of gaussian
  fits to helium absorption lines; the diagonal line is the 1:1
  relationship.  The 
$\lambda$4922 line widths measured here are systematically
  smaller than those reported by S17.}
\label{fig2}
\end{figure}


\begin{figure*}
\includegraphics[scale=0.6,angle=-90]{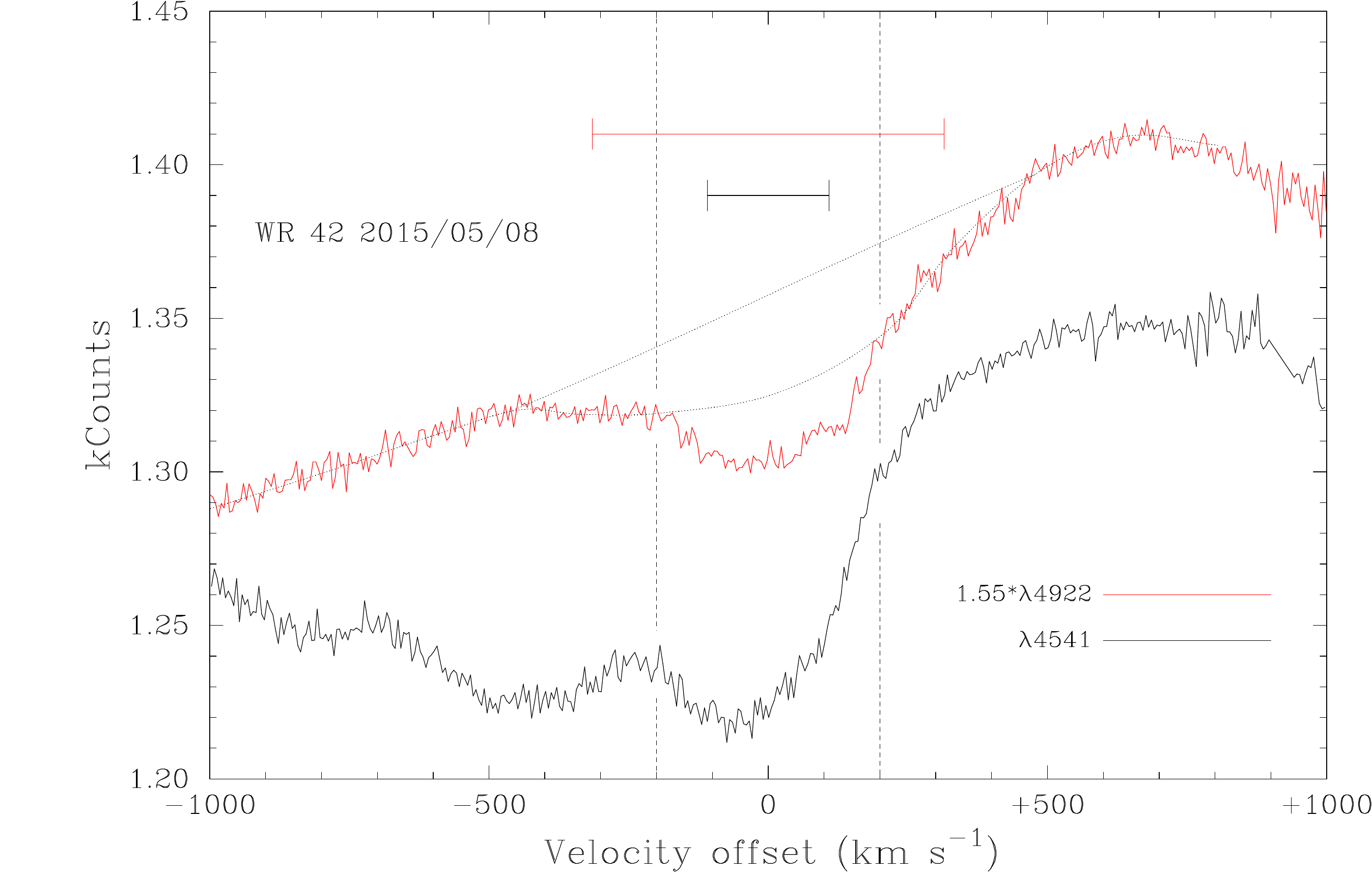}
\caption[FWHM plot3]{The \hefive, \henine\ lines in one of S17's
  observations of WR\;42, in velocity space. The $\lambda$4922 counts
  have been multiplied $\times1.55$ for display, and a small, ad hoc,
  global velocity shift has been applied to bring the absorption lines
  close to zero velocity.  Vertical dashed lines at $\pm$200~kms\ are
  intended only to facilitate comparison, and have no physical
  significance.  Dotted lines, discussed in $\S$\ref{sec:disco}, show
  two possible interpretations of the pseudo-continuum level
  appropriate to $\lambda$4922.  The horizontal bars represent the
  $\pm$\vmsini\ values found for each line by S17, demonstrating the
  factor $\sim$two difference they inferred for these lines from this
  spectrum.}
\label{fig3}
\end{figure*}

\subsection{Parameters}
\label{sec:par}

Given the abundances, microturbulence, and input physics, standard
model spectra are fully specified by two parameters describing the
atmosphere (normally \teff\ and \logg) and one describing the rotation
(normally \vesini, the maximum magnitude of the projection of the equatorial
rotation velocity onto the line of sight).  
For a gravity-darkened, rotationally distorted model star we may
equivalently specify
the corresponding global effective temperature,
\begin{align*}
\teff = \sqrt[4]{{\int{\sigma(\teffl)^4\,\text{d}A}}\left/{{{\int{\sigma\,\text{d}A}}}}\right.}
\end{align*}
(where $\sigma$ is the Stefan--Boltzmann constant and the integrations
are over surface area) and
the base-10 logarithm of the polar gravity in
c.g.s. units, \loggp.
However, we additionally require \textit{three} rotational
parameters because, for a rotationally distorted star, the equatorial
rotation speed and the axial inclination become separable, while the
magnitude of the gravity darkening depends on \omomc, the ratio of the 
rotational angular velocity\footnote{Assumed to be independent of latitude in
  the models discussed here.} to the critical value (Eq.~\ref{eq:vcrit}).

The physical parameters of the O-star components in the WR binaries
studied by S17 are poorly determined; in most cases, even the spectral
types are only approximate.  Rather than pursue `custom'
models, we therefore generated a grid of synthetic spectra to explore the parameter
space of interest.

The spectral types compiled by S17 for the O-star companions in their
sample range O4--O6 to O8--O9\,IV, with near-main-sequence luminosity
classes.  On that basis, we ran two series of models approximating
main-sequence and giant stars, adopting the parameters summarized in
Table~\ref{tab:models}.  The dependences on effective temperature of
polar gravity, \loggp, and polar radius, \rpole, are rough
approximations guided by the \citet{Martins05} calibration of O-star
parameters as a function of spectral type (their Tables~1 and~2).
The precise choices for these parameters are not critical;
ratios of line widths
are fixed for given \loggp, \teff, \omomc, and $i$ (although the overall
scaling of the system -- and hence the equatorial rotation velocity --
scales linearly with \rpole).

\begin{table}
\caption{Summary of model grids ($\S$\ref{sec:par});
note that \teff\ is in units of kK throughout this Table.
}
\begin{center}
\begin{tabular}{lllllllllllllllll}
\hline
\hline
Parameter && Range & Interval & Unit\\
\hline
\teff\   &    & 32:42 & 1   &kK\\
$\cos{i}$&    &0:1    & 0.1 &{\quad}--\\
\multicolumn{2}{l}{$\log_{10}(1 - \omomc)$}& $-$2:0 & 0.2 &{\quad}--\\
\loggp\  &(V) & 3.92  &     &dex cgs   \rule{0pt}{4ex}\\
&(III)& \multicolumn{3}{l}{$3.70 - 0.016\times(40 - \teff)$} \\
\rpole   &(V) &
\multicolumn{2}{l}{$(0.4\teff - 5)$}&\refrsun\\
&(III)& \multicolumn{3}{l}{$15.5 - 0.2\times(40 - \teff)$}\\
\hline
\end{tabular}
\end{center}
\label{tab:models}
\end{table}

\section{Results}
\label{sec:res}

The procedure adopted by \citeauthor{Shara17} (S17) was to rectify their
spectra using low-order polynomial fits to the continua in the region
of features of interest, followed by  least-squares gaussian fits to
characterize the full width at half maximum depth (FWHM) of the absorption
lines.  These FWHM values were then converted to measures of
rotational speeds  by using the polynomial
FWHM--\vesini\ relationships published by \citet{Ramirez15}.  

[Consequently, S17 tabulate their velocity measure\-ments as `\vesini' values.
However, given the systematic differences between results from
  \hei\ and \heii\ lines they report, these measures are clearly not
  intended to be interpreted as actual projected equatorial rotation
  speeds.  To avoid potential confusion, we will refer to these
  interpreted quantities as \vmsini\ (where the `m' subscript may be
  taken to indicate `measured'), reserving \vesini\ for the true
  projected equatorial rotation speed.]

  To characterize the model results in a manner as similar as possible
  to the observational results presented by S17 (and by
  \citealt{Ramirez15}) we simply fitted gaussians (plus a constant) to
  the helium lines of interest in the model spectra, following
  rectification with matched continuum models.  Particularly at high
  rotation speeds the lines can be shallow as well as broad, so in
  order to eliminate `wild' solutions (normally arising from blending
  with very weak helium) fits were rejected which yielded normalised
  central line depths of greater than 0.99 or central wavelengths more
  the 1\AA\ from the laboratory value.  For \henine, this limited the
  models to $\teff \leq 38$~kK.

We can circumvent issues associated with the inter\-mediate calibrations of \vesini\ as a
function of line width, and thereby more easily scrutinize the S17
line-width measure\-ments, by considering directly the helium-line full widths at
half maximum depth. The only
\heii\ line calibrated by \citeauthor{Ramirez15} is \hefive, while of the
\hei\ lines they considered only $\lambda$4922 is reasonably
straightfoward to measure in most of the S17 spectra.  Consequently,
S17 concentrated on the $\lambda\lambda$4541, 4922 lines -- as shall we.

%
%
%

 Model results for these lines are plotted in
Fig.~\ref{fig1}, along with measurements reported by S17 and
by \citeauthor{Ramirez15} (\citeyear{Ramirez15}; single stars, and
primary components of binaries).  We also include measure\-ments for the
late-O main-sequence stars HD~93521 and HD~149757 ($\zeta$~Oph),
obtained from the echelle spectra presented by \citet{Howarth01};
these are among the most rapidly rotating stars known
($\vesini \gtrsim 400 \kms$), and are
believed to have $\omomc \gtrsim 0.9$.

\section{Discussion}
\label{sec:disco}

It is apparent from Fig.~\ref{fig1} that the models are in broad
agreement both with the extensive
\citeauthor{Ramirez15} results, and with 
observations of the well-established
rapid rotators HD~93521 and $\zeta$~Oph.  The S17 measure\-ments,
however, are mostly offset to larger values of the
$\lambda\lambda$4922/4541 line-width ratio than either the models or
the bulk of
other observations.

While it is possible that this circumstance arises because the models
omit some relevant physics, or that the O stars in WR~binaries occupy
a region of parameter space not populated by other results, Fig.~1
suggest a more \mbox{prosaic} alternative -- that the S17
measure\-ments of the helium absorption lines may not all be reliable.
This would be perfectly understandable: the absorption lines are wide
and shallow (being both rotationally broadened and diluted by emission from the
companion), and are normally set within strong WR emission lines that
are likely to give rise to relatively steep and structured
pseudo-continua, with associated challenges to rectification.

To explore this possibility, we have carried out independent measure\-ments of the line
widths in the SALT spectra used by S17,\footnote{The reduced spectra
originally used by S17 have been mislaid; we are very grateful to Steve
  Crawford for providing re-reduced data to us.} following their
procedures except that, instead of approximating pseudo-continua by
low-order polynomials, we fitted Hermite splines to continuum points
selected by eye, which affords rather more flexibility in
accommodating the WR emission-line structure. Results are summarized in
Fig.~\ref{fig2}. While our measure\-ments of $\lambda$4541 are in
general agreement with S17's, our 
$\lambda$4922 FWHM values
are systematically smaller, by up to almost a factor~two.

Figure~\ref{fig3} illustrates the probable cause for these differences,
using observations of one of the most discrepant cases, WR~42. 
The figure emphasizes the
importance of continuum placement for these shallow absorption
features (typical depths 2--3\%\ of local pseudocontinuum levels).  In
this case a `high' continuum for $\lambda$4922 was reconstructed by
dividing the observed spectrum by the S17 gaussian fit, and 
consequently should
be
a reasonably close match to their choice.  Our alternative `low'
continuum is, we suggest, at least equally plausible from a purely empirical
perspective, and leads to a line width that is, in practice,
indistinguishable from that for $\lambda$4541.  Thus, while there is no
fully objective way of deciding which (if either) of the proposed
continua is `correct', we believe that Fig.~\ref{fig3} demonstrates that
exceptionally strong gravity-darkening effects are not necessarily required in
order to
explain the observations; a conservative interpretation of the results
is therefore that they are consistent with model-based expectations.

\begin{figure}
\includegraphics[scale=0.6,angle=-90]{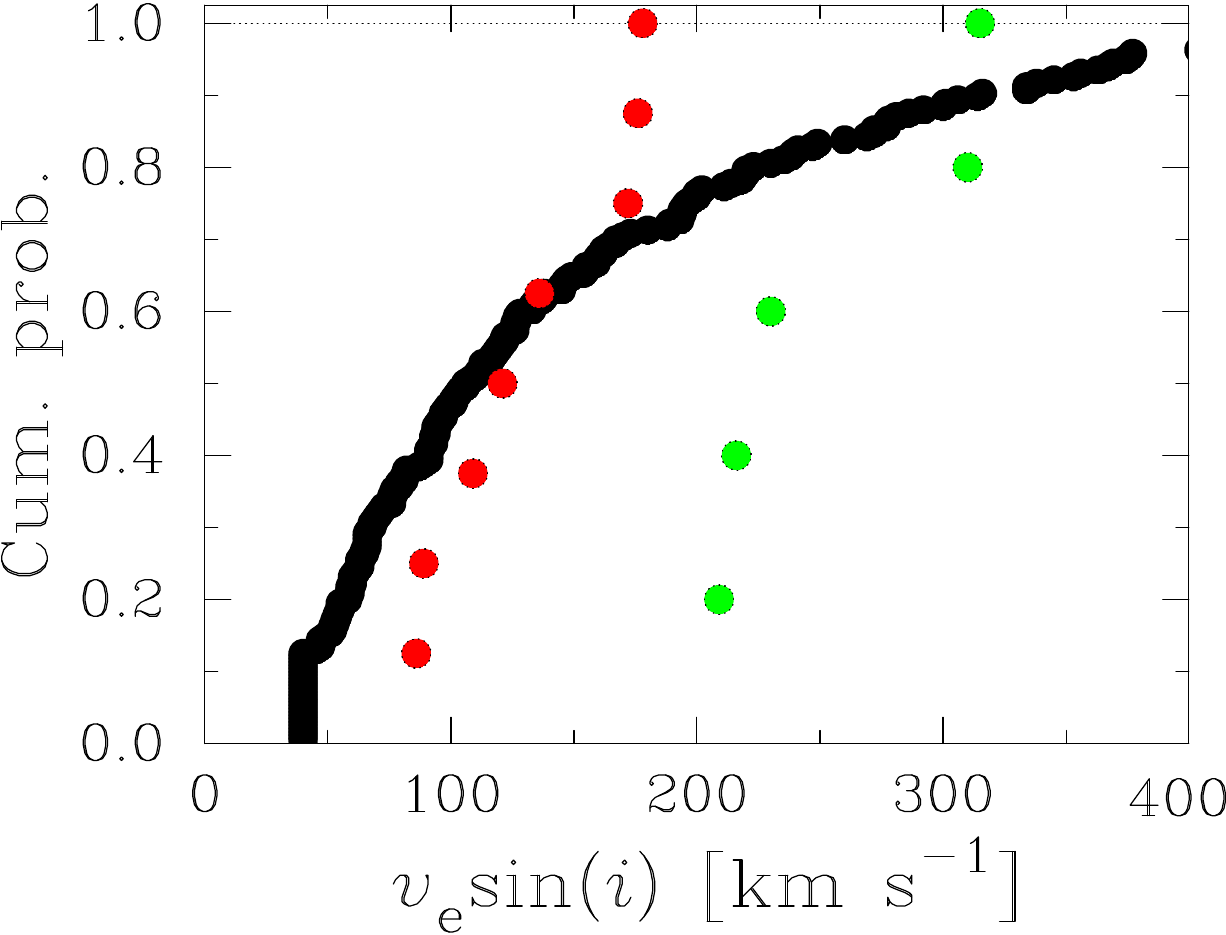}
\caption[vesini plot4]{Cumulative probability distribution functions
  of projected rotational velocities.  Black dots: \vesini\ measures
  for single O stars, from \citeauthor{Ramirez13} 
(\citeyear{Ramirez13}; the cut-off at $\vesini = 40$~\kms\ is
observational, not astrophysical).  Red [green] dots:
  \hefive\
  [\henine]
  \vmsini\ measures for O stars in WR binaries (from S17);
where multiple
  measure\-ments are available for a given system, the value with the
  smallest error was used.}
\label{fig4}
\end{figure}

A consequence of this is that the \vesini\ value for each of the
O-star components could be at the lower (\hefive) end of the \vmsini\
values reported by S17, rather than the high-end (\henine) values they adopt.
In that case, based on the synchronous-rotation rates compiled by S17,
most -- though now not all -- of the O-star rotation speeds remain
supersynchronous.  However, it is unclear if this requires any special
spin-up mechanism, as suggested by S17.  Fig.~\ref{fig4}
compares the cumulative probability distribution functions of inferred
rotational velocities for the S17 sample (\vmsini\ values) with the
\vesini\ measure\-ments reported by \citet{Ramirez13} for a sample of
apparently single O~stars.  If we adopt the \hefive\ \vmsini\ values
as more representative of the projected equatorial velocities than are
the \henine\ values, then it appears plausible that the
supersynchronous rotation in wide binaries could arise simply through
initial conditions that are unexceptional -- in fact, it is the
\textit{absence} of very rapid (and very slow) rotators that stands
out in Fig.~\ref{fig4}.

A Kuiper test confirms the qualitative impression that, even for the
small-number statistics that apply here, the null hypothesis that CDFs
for the single and S17 O stars are drawn from the same parent
populations can be rejected with $\sim$99\%\ confidence.  Of course,
the comparison made in Fig.~\ref{fig4} is subject to many caveats, and
the S17 and \citeauthor{Ramirez13} samples are, in several respects,
not directly comparable; but again, a conservative interpretation
allows for the possibility that there is no strong \textit{a priori}
case for suggesting that the WR+O systems require special
consideration in the context of current tidal-braking theory
(nowithstanding its other shortcomings; cf., e.g.,
\citealt{Khaliullin10}).

\section{Conclusions}

We have re-examined the rotational velocities of O stars in WR+O
binaries.  New model calculations and analyses
of large samples of `normal' stars
are in good mutual agreement, but published measure\-ments of the
WR+O systems are discrepant with both.  We have shown that this
discrepancy can reasonably be explained by the choice of pseudo-continuum
levels, particularly for the shallow \henine\ line.   Consequently, we
suggest that the observations demand neither implausibly large
gravity-darkening effects, nor novel mechanisms to sustain
supersynchronous rotation.

\section*{acknowledgements}

We are especially grateful to Steve Crawford and Mike Shara for
generously providing a complete set of the SALT spectra used by
S17, and for encouragement in this study.  Oscar Ramirez and Hugues
Sana kindly supplied the unpublished FWHM measure\-ments that underpin
the \citet{Ramirez15} calibrations.  We also thank Ivan Hubeny for provision of
\textsc{tlusty} and associated programs, and for support during
the calculation of our intensity grids.

\bibliographystyle{mnras}
\bibliography{IDH}

\label{lastpage}

\end{document}